\documentclass[smallextended]{svjour3}     
\usepackage{graphicx}
\usepackage{mathptmx}    
\journalname{Gen. Rel. Grav.}

\begin{document}

\title{Cosmological background solutions and\\
cosmological backreactions}

\titlerunning{Background Soultions and Backreactions}

\author{Edward W.\ Kolb \and Valerio Marra \and Sabino Matarrese}

\authorrunning{Kolb \and Marra \and Matarrese}

\institute{
Edward W.\ Kolb \at
Department of Astronomy and Astrophysics, Enrico Fermi Institute,
and  Kavli Institute for Cosmological Physics, The University of
Chicago, Chicago, IL 60637-1433, USA\\
\email{rocky.kolb@uchicago.edu}
\and
Valerio Marra \at
Department of Physics, University of Jyv\"{a}skyl\"{a}, PL 35 (YFL), 
FIN-40014 Jyv\"{a}skyl\"{a}, Finland
and Helsinki Institute of Physics, University of Helsinki, PL 64, 
FIN-00014 Helsinki, Finland\\
\email{valerio.marra@jyu.fi}    
\and
Sabino Matarrese \at
Dipartimento di Fisica ``G.\ Galilei'', Universit\`{a} di Padova, 
INFN Sezione di Padova, via Marzolo 8, Padova I-35131, Italy\\
\email{sabino.matarrese@pd.infn.it}    
}

\date{Received: date / Accepted: date}

\maketitle

\begin{abstract}

The cosmological backreaction proposal, which attempts to account for
observations without a primary dark energy source in the stress-energy tensor,
has been developed and discussed by means of different approaches. Here, we
focus on the concept of cosmological background solutions in order to develop a
framework to study different backreaction proposals.



\end{abstract}

\section{Backgrounds and backreactions}

A cosmological background solution is a mean-field geometry of suitably averaged
Einstein equations, in which the average expansion is described by
a single scale factor.
A cosmological background solution depends upon the spatial curvature and the mass-energy content.
Associated with the latter is a stress-energy tensor that describes a
fluid (or fluids) with equation(s) of state that satisfy local energy
conditions.
For simplicity, we will assume that there is only one fluid, with a
local equation of state of pressureless matter.

If the universe were exactly homogeneous and isotropic, there would be a
unique background solution that describes the expansion history of every volume
element as well as observations made by any observer.
In particular the background solution would be a Friedmann-Lema\^{i}tre-Robertson-Walker (FLRW) solution of the Einstein
equations.

In this paper we consider the more relevant case of an inhomogeneous universe,
and consider the description of the evolution of such a universe by  perturbing
about some cosmological background solution.  The issue is which cosmological
background solution to perturb about to best describe the expansion history and
observables on our past light cone.  This paper is a critical study of these
issues.  

We start by defining and elucidating several types of cosmological background
solutions focusing on energy/curvature content, dynamics and observations:
\footnote{New definitions will be marked in {\bf bold type} throughout the paper.}

{\bf Global Background Solution (GBS)}: a background solution with energy
content and spatial curvature that are the (particle) horizon-volume spatial averages of
the energy content and the spatial curvature of the inhomogeneous universe. 
The other assumption is that the fluid equation of state is the local equation
of state of the fluid comprising the stress tensor.
The GBS is therefore the usual FLRW background, in particular it is not a function
of position.

{\bf Averaged Background Solution (ABS)}: a background solution that describes
the evolution of the horizon volume of an inhomogeneous universe.  The energy
content and the curvature that generate the ABS are not necessarily related to
the spatial averages of the energy content and curvature of the spatial region
considered. In particular, the equation of state of the stress-energy tensor of the
ABS, even though is supposed to satisfy the dominant energy condition,
need not be related to the local one.
The ABS may be obtained by averaging over the horizon
volume following, for example, the formalism introduced by Buchert
\cite{Buchert:2007ik}. The ABS is not a function of
position.

{\bf Phenomenological Background Solution (PBS)}: a background solution that
effectively describes the observations, that is, the data on the past light cone
of an observer. The energy content and curvature of the PBS are not necessarily
related to the spatial averages of the energy content and curvature of the
observable universe. The equation of state of the stress-energy tensor of the
PBS need not satisfy any of the local energy conditions of the mass-energy
content of the universe.
The relevant PBS may depend on the position of the observer and the redshift out
to which the observations are performed. 

The {\it standard} approach consists of assuming that the three background
solutions coincide in the sense that they are perturbatively close to one
other.  In this approach one describes the spacetime, the global dynamics, and
the observations by means of only one background solution, namely the GBS. Our
purpose here is a critical re-examination of this simplification.

The next step is to define what we mean by backreaction:

{\bf Backreaction}: the situation in which the three backgrounds defined above
{\it do not} coincide. 

We then have two possibilities for the backreaction:

{\bf Strong Backreaction}: the GBS does not describe the expansion history, {\it
i.e.,} the GBS and the ABS do not coincide.  The ABS and the PBS may,
or may not, coincide.

{\bf Weak Backreaction}: the GBS coincides with the ABS. The
PBS, however, differs from the GBS.

Summarizing, a strong backreaction deals with the non-commutativity between averaging
and going to the field equations (EoM), while a weak backreaction deals with the non-commutativity
between averaging and measuring (Obs). Epigrammatically:
\begin{eqnarray}
\left [ \langle\cdots\rangle, \, \textrm{EoM} \right] \neq 0 \, &\Rightarrow& \,
\textrm{ABS} \neq \textrm{GBS} \nonumber \\
\left [ \langle\cdots\rangle, \, \textrm{Obs} \right] \neq 0 \, &\Rightarrow& \,
\textrm{PBS} \neq \textrm{GBS} \nonumber
\end{eqnarray}
%

\section{The Global Background Solution and the FLRW Assumption}

In cosmology, one usually models the evolution and observables associated with
an inhomogeneous universe consisting of density $\rho(\vec{x})$,
three-curvature $^3{\cal R}(\vec{x})$, and expansion rate $H(\vec{x})$ by
employing a FLRW model of density $\rho = \langle\rho(\vec{x})\rangle$, and
spatial curvature $^3{\cal R}=\langle^3{\cal R}\rangle$ (where
$\langle\cdots\rangle$ denotes some suitably defined spatial average) and a
fluid equation of state given by the local equation of state of the fluid(s)
comprising the stress-energy tensor.  One then assumes that the expansion
history and cosmological observables are those obtained in the corresponding
FLRW model (the GBS). Inhomogeneities are thought of as developing in this
fixed GBS according to perturbation theory.  In other words, it is implicitely
assumed that the GBS is the only relevant background solution and there are no
backreactions.

In this approach, inhomogeneities are, therefore, simply neglected: one assumes
that inhomogeneities will not affect this long-standing procedure without
actually providing a quantitative averaging method to demonstrate it. This
approach does not provide any guidance as to the scale upon which such models
are supposed to be applicable, nor does it seriously examine the issues
arising when we consider the relations between descriptions of the universe at
different scales of inhomogeneity \cite{ellis,Carfora:1984,Ellis:1987zz,Ellis:2008zz}.

It is usually stated that the above approach is justified by the cosmological
principle, which simply states that the universe is homogeneous and isotropic
on a sufficiently large scale.   However, there is an unstated {\it assumption}
implicit in addition to the cosmological principle. We first state the
assumption:

{\bf  FLRW Assumption:} in an inhomogeneous universe that satisfies the
cosmological principle, the volume expansion and cosmological observables may
be described by perturbations of the GBS.  

The FLRW assumption, of course, implies the absence of any backreaction.
In support of the FLRW assumption, it has been claimed
\cite{Ishibashi:2005sj,Gruzinov:2006nk} that even in the presence of highly
nonlinear density perturbations ($\delta\rho/\rho\gg1$) the metric for our
universe can everywhere be written as a perturbed conformal Newtonian metric of
the form  
\begin{equation} 
ds^2=a^2(\tau)\left[-(1+2\psi)d\tau^2 + (1-2\psi)\gamma_{ij} dx^idx^j \right]  ,
\label{newtpert} 
\end{equation} 
where $d\tau = dt/a$ is conformal time, $\gamma_{ij}$ is a metric of a
three-space of constant curvature, and  $\psi$ satisfies the Newtonian
conditions $|\psi| \ll 1$,  $\left| \partial\psi/\partial t \right|^2 \ll 
a^{-2}D^i\psi D_i\psi$, and $\left(D^i\psi D_j\psi\right)^2\ll
\left(D^iD^j\psi\right)D_iD_j\psi$.  The covariant derivative with the metric
$\gamma_{ij}$ is denoted by $D_i$.  The usual statement is that in the dust
case one is allowed to use the perturbed conformal Newtonian metric either in
the linear regime (\textit{i.e.,} perturbations of every quantity being small)
or in the weak-field (Newtonian) regime.\footnote{But note that already at
second  order in perturbation theory the spatial part of the metric differs
from the  temporal part of the metric simply from terms quadratic in the
peculiar velocities (\textit{i.e.,} an effective anisotropic stress term is in
the stress-energy tensor).} In other words, according to this, the GBS
describes the evolution of an inhomogeneous universe even  in the presence of
large inhomogeneities, and therefore there is no strong backreaction. The
latter claim has indeed been used as a ``no-go'' theorem against the
backreaction proposal.

In the following we consider the case of a (on average) spatially-flat
universe, consisting only of dust, with no primary dark energy.  The GBS is,
therefore, the Einstein-de Sitter (EdS) solution.  According to the FLRW
assumption, strengthened by the no-go argument, the EdS solution describes the
global expansion and the observables.   If there would be a backreaction, the
EdS global background solution would not describe the expansion or the
observables.

\section{Peculiar velocities}

In Ref.\ \cite{Kolb:2008bn} we explored assumptions that could be relaxed  in
order to evade the aforementioned no-go theorem.  In that work we concentrated
on one of the criteria that defines the Newtonian  regime, namely ``velocities
much smaller than light relative to the Hubble flow'' \cite{Ishibashi:2005sj}.
To appreciate the subtilties in this criterion, we first must refine the
concept of peculiar velocities.  

A ``peculiar'' velocity is always defined with respect to the velocity of a
background solution, \textit{i.e.,} a background Hubble flow.   The subtilty is
``which'' Hubble flow.
We start by defining different peculiar velocities depending on which
background is chosen as a reference:

{\bf Global Peculiar Velocities}: velocities obtained after subtracting the
Hubble flow of the GBS.

{\bf Averaged Peculiar Velocities}: velocities obtained after subtracting the
Hubble flow of the ABS.

{\bf Phenomenological Peculiar Velocities}: velocities obtained after
subtracting the Hubble flow of the PBS.

Phenomenological Peculiar Velocities are departures from the observed Hubble
flow predicted by the
Phenomenological Background Solution. Observations tell us that
Phenomenological Peculiar Velocities are small (in the sense of being much less
than $c$).

Global Peculiar Velocities, instead, are velocities relative to the Global
Background Solution, and have nothing to do with anything that can be measured
as a local effect. Of course, in the usual approach the Phenomenological
Background Solution coincides with the Global Background Solution, and
therefore, Phenomenological Peculiar Velocities and Global Peculiar Velocities
coincide.

In Ref.\ \cite{Kolb:2008bn} we found that one way to evade the no-go argument
is to relax the restriction that the global peculiar velocities must
be small, while still demanding small phenomenological peculiar velocities. 
Small phenomenological peculiar velocities can be seen as a requirement
coming from observational data.
We see, therefore, that the FLRW assumption asks for small global
peculiar velocities. We will see that this is a
restriction on the dynamics of inhomogeneities, in particular, voids.

\begin{figure*}
\includegraphics[width=0.85\textwidth]{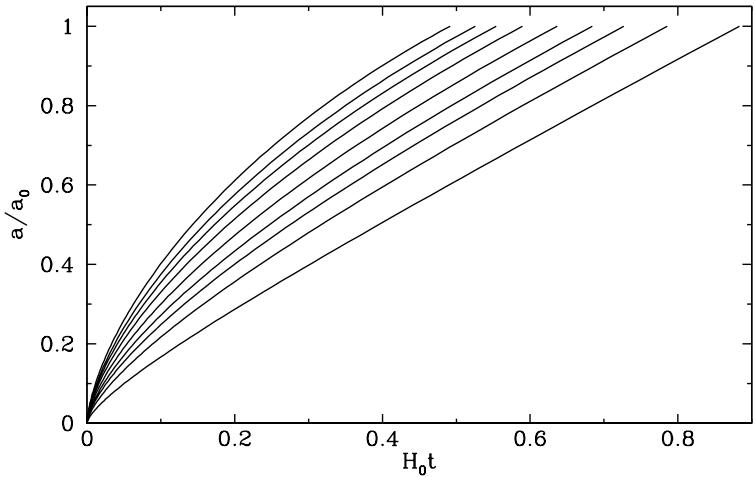}
\caption{The time evolution of the scale factor for different FLRW solutions.}
\label{FLRWs}
\end{figure*}

We should, indeed, regard inhomogeneous regions as different FLRW models
characterized by a different local density and curvature, that is, as regions
with different and differently evolving expansion rates. If we do not restrict
the different patches to evolve close to the GBS, they will naturally develop
large global peculiar velocities because different FLRW solutions follow
different evolutions as sketched in Fig.\ \ref{FLRWs}, where we have plotted
the evolution of the scale factor for FLRW models with different curvature. It
can happen that at some specific time, due to initial conditions, all the
inhomogeneous regions are close to some GBS. This will be, however, a special
moment because of the different evolution of the patches. It will be more
typical to have large departures from the GBS Hubble flow, which will be
manifest in a large $\Delta H(\bf {x})$ or in large global peculiar velocities
that are the results of perturbing about the wrong background. The relevance of
the GBS may, therefore, be more fragile than previously appreciated, while the
relevance of PBS seems to be more robust (as can be seen by the success of
$\Lambda$CDM).

\section{Observers}

In the previous two Sections we focused on the influence of inhomogeneities in
the description of the spacetime of the universe. In this Section we now turn
to discuss the possibility that the GBS/ABS is not guaranteed to be able to
describe observations. This discussion applies to both the weak backreaction,
in which case GBS/ABS will mean GBS, and the strong backreaction, in which case
GBS/ABS will mean ABS because in this case the GBS does not describe any aspect
of the universe.
We start by defining two types of observers.

{\bf Global Observer}: is an observer comoving with the GBS/ABS Hubble flow. 

{\bf Phenomenological Observer}: is an observer comoving with the PBS Hubble
flow, that is, after subtracting any local peculiar velocity. This observer
measures the phenomenological background which is the only background directly related
to observations.

The cosmological principle states that the universe is homogeneous and
isotropic on a sufficiently large scale. As pointed out above, the cosmological
principle, however, does not tell us how the large-scale homogeneity and
isotropy are connected to the description of the universe at smaller scales. In
particular, it does not deal with the observer: it does not follow from the
cosmological principle that every phenomenological observer is equivalent
independent of location, neither that the global observer has anything to
do with the phenomenological observers.

The usual approach is the following. The Copernican principle states that
phenomenological observers do not occupy a privileged position in the universe.
From the Copernican principle, together with the observed isotropy of the universe (from
the CMB, for example), the cosmological principle follows. One then makes use
of the FLRW assumption, and every observation can be described by the GBS. The
success of the concordance model then verifies this reasoning {\it a
posteriori}.

However, the only truly {\it direct} consequence of the success of the
concordance model is that the isotropic and homogeneous $\Lambda$CDM model is a
good  {\it phenomenological} fit to the real inhomogeneous universe, but this
does not imply that a primary source of dark energy  exists, but only that it
exists as far as the phenomenological fit is concerned. For example, it is not
straightforward that the universe is locally accelerating in the usual sense.

We therefore reconsider the latter approach in order to have principles closer
to the ``unprocessed'' observations. We choose the cosmological principle as
more fundamental than the Copernican principle, and we define a new type of
cosmological principle:

{\bf Bare Cosmological Principle:} the universe is homogeneous and isotropic on
a sufficiently large scale. 

Equivalently, the bare cosmological principle states that it is possible to
describe the universe by means of a mean-field approach, that is, within our
definitions, that the Averaged Background Solution becomes, at the horizon
scale, insensitive to the scale of averaging: in other words, it is possible to
describe the inhomogeneous universe by means of a suitable cosmological background solution (but not
necessarily the FLRW/GBS).

From the observed isotropy (in agreement with the bare cosmological principle)
together with the observational success of the concordance model we can then
state a new Copernican Principle:

{\bf Bare Copernican Principle:}  every observer can describe the universe by
means of a mean-field description, that is, it is possible to describe
observations in the inhomogeneous universe by means of a suitable spherically-symmetric
cosmological background solution
(but not necessarily the GBS/ABS) which is the Phenomenological Background
Solution, even though differently located observers may have different PBSs. 

We have therefore at least two different situations where 1) every
phenomenological observer sees the same PBS and 2) every phenomenological
observer sees a different PBS.

\section{Examples and discussion}

In this Section we will discuss the concepts introduced so far by means of concrete examples.
It will be clarifying to consider, among others, swiss-cheese models made of dust, which we now briefly introduce.

By swiss-cheese model we will mean the general setup in which
spherically symmetric Lema\^{\i}tre-Tolman-Bondi (LTB) metrics \cite{Lemaitre:1933gd,Tolman:1934za,Bondi:1947av}
(the holes) are embedded in the EdS metric (the cheese).
Assuming a dust equation of state, the Einstein equations for the LTB model can
be solved to obtain the dynamical equation
\begin{equation}
\label{Hb}
\frac{\dot{a}^{2}(r,t)}{a^{2}(r,t)} = \frac{8 \pi G}{3}\, 
\hat{\rho}(r,t)  - \frac{k(r)}{a^{2}(r,t)} ,
\end{equation}
where $\hat{\rho}(r,t)$ is the (euclidean) averaged density up to the shell $r$ and
the curvature term $k(r)$ can be interpreted as the total energy of the shell,
which is given by its kinetic energy plus its potential energy due to the total
(euclidean) mass up to the shell $r$ (see Ref.\ \cite{Bondi:1947av} from more
details).

Equation (\ref{Hb}) is the generalization of the Friedmann equation for a
homogeneous and isotropic universe to a spherically symmetric inhomogeneous
universe. To recover the usual Friedmann equation we replace $a(r,t), \,
\hat{\rho}(r,t), \, k(r)$ by $a(t), \, \rho(t), \, \pm1$ or $0$.
From Eq.\ (\ref{Hb}) it follows that every shell $\bar{r}$ can be thought as a
different FLRW universe of density $\hat{\rho}(\bar{r},t)$ and curvature
$k(\bar{r})$.
Indeed, if the LTB model does not undergo shell crossing (see Ref.~\cite{Hellaby:1985zz}),
each shell evolves independently of the others.
Spherical symmetry severely restricts the dynamics of the shells. Equation
(\ref{Hb}) shows that, in a spherically symmetric system,
the dynamics of the shell $r$ depends only on the total mass within and does
not depend on the mass outside. This clearly shows how density inhomogeneities
are averaged out with respect to the dynamics of the shells.

\begin{figure*}
\includegraphics[width=0.6\textwidth]{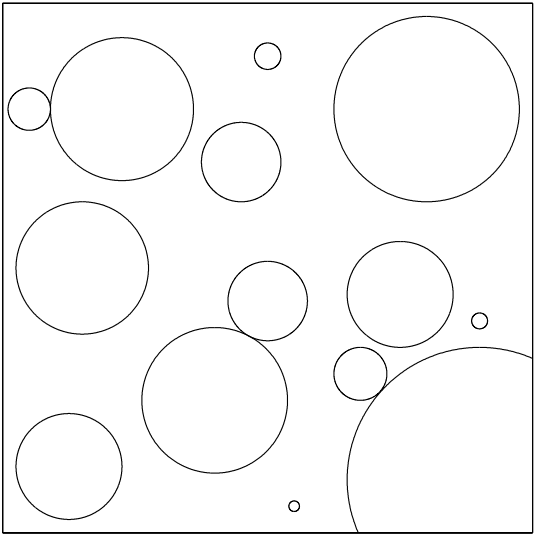}
\caption{Sketch of an exact swiss-cheese model.}
\label{sc}
\end{figure*}

In order to have an exact swiss cheese it is necessary to avoid any
superposition among the holes (as sketched in Fig.\ \ref{sc}). Moreover, the
LTB solution of each hole has to match the EdS metric of the cheese at the
boundary of the hole at $r=r_{h}$. This means that
$\hat{\rho}(r_{h},t)=\rho(t)$ and $k(r_{h})=0$ as it is clear from Eq.\
(\ref{Hb}). As for the density, this implies that if there is an underdensity,
there must be a compensating overdensity (see Ref.\ \cite{Marra:2007pm,Marra:2007gc} for more details).

Thanks to the matching, an observer outside the holes will not feel the
presence of the holes as far as \textit{local} physics is concerned (this does
not apply to global quantities, such the luminosity-distance--redshift relation
for example). So the cheese is evolving as an EdS universe while the holes
evolve differently. Independent of how many holes we put in, we will still have
an exact solution of the Einstein equations. Again, this is due to spherical symmetry.
We can imagine placing holes in order to satisfy the bare cosmological
principle.
We therefore see that, {\it by construction}, in an exact swiss-cheese model
the scale factor of the ABS evolves as the one of the EdS model, which describes the evolution of the
swiss-cheese on scales larger than any spherical hole. In other words, in this
special case, a cheese-only model and the swiss-cheese model have the same
volume evolution in spite of the inhomogeneities. This tells us that the
swiss-cheese model is just by construction the wrong model to study the impact of
inhomogeneities on the ABS, that is, to study strong backreaction.

The swiss-cheese model is, however, a useful toy model with which to study
the GBS and the PBS. In the following we will discuss
some swiss-cheese models together with other examples found in the literature. 
As we will see, the
backreaction proposal can act through different ways. We would like to stress
that a combination of the possible effects might be necessary to explain away
dark energy.

\subsection{The GBS does not describe the spacetime}

As we have seen, independently of the inhomogeneities, in swiss-cheese models the ABS evolves as the EdS model.
Moreover, the curvature inside the holes is small (quadratic in hole-radius/Hubble-radius) as far as metric quantities are concerned, and so EdS and
swiss cheese share almost the same average density.
It appears, therefore, natural to perturbe around the EdS model and identify it as the GBS.
Yet, until a suitable averaging procedure is carried out (see, for example, Ref.~\cite{Buchert:2002ht}), it is not possible to state that
the GBS, within our definitions, is the EdS.
It is not excluded that the actual GBS for a swiss-cheese model is not a perturbation of EdS.
However, in the spirit of the no-go argument that states the ABS and GBS coincide, it is reasonable to choose the EdS model as the background to perturb from in eq.~(\ref{newtpert}).

In Ref.\ \cite{Kolb:2008bn}, we found an example of a model \cite{Alnes:2005rw}
that fits the luminosity distance and the position of the first CMB peak, but
evades the no-go argument. 
The observer is located at the center of a Gpc-scale underdensity which mimics a temporal variation of the Hubble parameter.
We found large global peculiar velocities, which
can be generally expressed as:
\begin{equation}   
\label{pecp}  
v_\textrm{\scriptsize GPV} \simeq   R \cdot H_\textrm{\scriptsize GBS} \cdot  
\frac{\Delta H}{H_\textrm{\scriptsize GBS}} \,. 
\end{equation}  
This can be understood to show that to the global peculiar velocities,
$v_\textrm{\scriptsize GPV}$, contribute the physical size of the
inhomogeneity, $R$, the magnitude of the global background Hubble parameter,
$H_\textrm{\scriptsize GBS}$, and the fractional departure of the expansion of
the inhomogeneous region from the global background Hubble parameter, $\Delta
H/H_\textrm{\scriptsize GBS}$.

We stress again the importance of not assuming {\it a priori} small global
peculiar velocities. Otherwise, we fall into a circular reasoning in which we
assume as a starting point that the GBS describes the spacetime, and hence the
absence of strong backreaction. If indeed inhomogeneities alone explain the
concordance model, then there will be large global peculiar velocities with
respect to the GBS/EdS-model which would be a wrong background.

\subsection{The PBS differs from the GBS}

In this Section we will consider three examples which will help understanding how the PBS can differ from the GBS: a swiss-cheese model, a meatball model and a two-scale model.

\begin{figure*}
\includegraphics[width=0.85\textwidth]{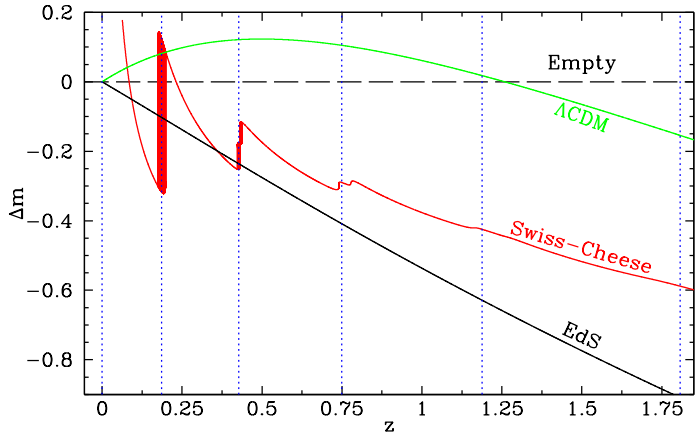}
\caption{The luminosity distance as a function of redshift for an observer looking through the holes of the swiss-cheese universe of Ref.\ \cite{Marra:2007pm,Marra:2007gc}, together with the GBS/EdS and $\Lambda$CDM curves. Rather than $H_0 d_L(z)$, we show the usual difference
in the distance modulus compared to the empty model.}
\label{obsincheese}
\end{figure*}

In Ref.\ \cite{Marra:2007pm,Marra:2007gc}, we have studied the propagation of
light in a swiss-cheese universe. In that case global peculiar velocities were
small enough to have a Newtonian description \cite{VanAcoleyen:2008cy}, that
is, the GBS/EdS-model describes perturbatively the spacetime.
We found, however, a sizeable weak backreaction caused by the evolution of
inhomogeneities, in particular voids. The setup was of an observer in the
cheese looking through a chain of holes. The observer was, therefore, a global
observer not in a special position, but the observables differed from those
calculated in the GBS as shown in Fig.~\ref{obsincheese}.

The effect of the inhomogeneities on the observables we have found can be imputed to lensing effects as discussed in Ref.\ \cite{Vanderveld:2008vi,Ghassemi:2009ug,Valkenburg:2009iw}.
It has been argued in Ref.\ \cite{VanAcoleyen:2008cy,Vanderveld:2008vi} that the results found in Ref.\ \cite{Marra:2007pm,Marra:2007gc} converge to the EdS result once an angular average is considered: in Ref.\ \cite{Marra:2007pm,Marra:2007gc} the observer was, indeed, only looking through the center of the holes.
However, the swiss cheese studied in Ref.\ \cite{Marra:2007pm,Marra:2007gc} is not suitable for statistical averaging.
Because of the setup a photon always hits a surrounding overdensity before and after passing through an underdensity: the lensing probability distribution function (PDF) is therefore bound to not give a large net effect as also shown by Ref.~\cite{Brouzakis:2007zi}. 
In order to study the PDF it is necessary to focus on a model that allows photons to miss overdensities, like, for example, the meatball model of Ref.~\cite{Kainulainen:2009sx,Kainulainen:2009dw} (see also the lattice model of Ref.~\cite{Clifton:2009jw}): in this case, even with randomly placed meatballs, an observer can indeed observe a PBS different from the GBS.
The advantage of swiss-cheese models based on LTB solutions is instead in their ability to describe the local spacetime.
We now see that the direction through voids of Ref.\ \cite{Marra:2007pm,Marra:2007gc} can be seen as a representative one: the photons we observe
have likely travelled more through voids than through structures (see also Ref.\ \cite{Kantowski:1969,DyerRoeder1,Mattsson:2007tj}).

The fractal bubble model of Ref.~\cite{Wiltshire:2007jk,Wiltshire:2007fg,Leith:2007ay} and the two-scale model of Ref.~\cite{Rasanen:2006kp} focus on the two relevant cosmological scales of the late universe: voids that dominate the volume and structures, in the form of filaments and walls, that surround the voids.
In contrast to swiss-cheese models, a compensation among perturbations is not demanded, the two scales evolve indeed independently.
Moreover, the fractal bubble model takes into account the differences in proper time between observers within voids and observers within structures with the result that the PBS (for an observer within a structure) is found different from the ABS.
In particular, the PBS exhibits apparent acceleration, while the ABS does not.
This is shown by means of a Buchert average, which also shows that PBS and ABS differ from FLRW solutions and in particular from the GBS.

\subsection{Different PBSs for different phenomenological observers}

In the previous section we have seen examples of models that feature a PBS different form the ABS/GBS.
We will now discuss two possible mechanisms that can make the PBS differ for different phenomenological observers.

The first concerns the dressing of cosmological parameters for observers using an FLRW bias to describe observations.
As explained by Ref.~\cite{Buchert:2002ht,Buchert:2002ij}, once we have Buchert averaged a given model, we still have to consider how the geometrical inhomogeneities dress the fiducial FLRW density parameters of volume effects due to the differences between the smoothed FLRW volume and the actual volume of the inhomogeneous region considered.
Furthermore, Ref.~\cite{Buchert:2002ht,Buchert:2002ij} found that the spatial curvature is also dressed by curvature backreaction effects which encode the deviation of the averaged scalar curvature from a constant curvature model.
The dressing of the parameters, therefore, not only makes the PBS depart from the ABS, but also links different PBSs to different phenomenological observers.

\begin{figure*}
\includegraphics[width=0.85\textwidth]{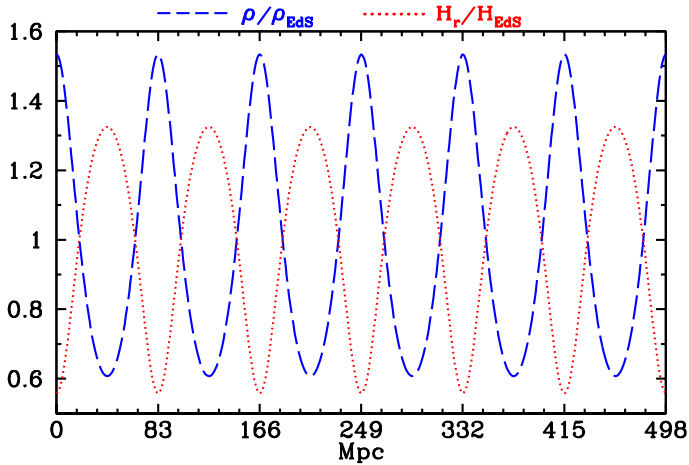}
\caption{Density profile (dashed blue line) and radial expansion rate (dotted
red line). The model exhibits a large variance around the average EdS values.}
\label{rho&Hr}
\end{figure*}

As a second example we will consider a specific LTB toy model with a sinusoidal density
profile\footnote{We present here a model similar to the one discussed in Ref.\ \cite{Marra:2008sy}, Chapter 3. More precisely, the initial density and curvature are $\rho(r,\bar{t})=\rho_\textrm{\scriptsize GBS}(\bar{t}) \left(1+0.3 \cos {2 \pi \over \lambda} r \right)$ and $E(r)=0.8 \, \left({1 \over 2}H_\textrm{\scriptsize GBS}(\bar{t})^{2}a(\bar{t})^{2}r^{2}-GM(r)/a(\bar{t})r \right)$. The initial time $\bar{t}$ corresponds to $z\simeq 1.9$ and $\lambda \simeq 83.5$Mpc. Moreover,  $M(r)={4 \pi \over 3}\hat{\rho}(r,\bar{t})a(\bar{t})^{3}r^{3}$ and $k(r)=-2E(r)/r^{2}$.} and we focus on a region of it far from the center.
This model belongs to the class of onion models \cite{Biswas:2006ub,Marra:2008sy} and,
similarly to swiss-cheese models,
is matched to the GBS/EdS-model in between every oscillations.
The size of the inhomogeneities is around $80$ Mpc and
the cosmological principle is respected. We show in Fig.\ \ref{rho&Hr} a plot
of the present-time density and radial expansion rate which is defined as
$H_{r}=\dot{R}'/R'$ where $R=r\,a(r,t)$.
In this model the GBS/EdS-model describes perturbatively the spacetime and
there are no sizeable effects on the propagation of light.
We would therefore expect that global and phenomenological observer coincide.

\begin{figure*}
\includegraphics[width=0.85\textwidth]{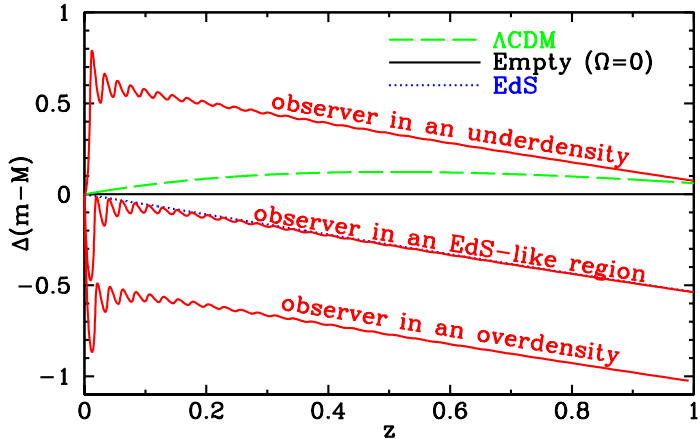}
\caption{The luminosity distance as a function of redshift for an observer in
an underdense region, in a region with parameter close to the GBS/EdS-model and
in an overdense region. Rather than $H_0 d_L(z)$, we show the usual difference
in the distance modulus compared to the empty model.}
\label{dm}
\end{figure*}

The point we want to make is that the global observer is meaningful, that is,
he says something about the phenomenological observer only if the
inhomogeneities are at the linear order. When they become nonlinear, the
phenomenological observer might be different from the global observer. In Fig.\
\ref{rho&Hr} we see indeed that the average EdS expansion rate does not say
anything about the local expansion rate because the variance is so large.
As a result, the luminosity distance is observer-dependent as shown in Fig.\
\ref{dm} where we compare the luminosity distance measured by an observer in an
underdense region, in an overdense region and in a region with parameter close
to the GBS/EdS-model. This scenario is usually named Hubble
bubble scenario in order to stress the fact that, in order to fit the luminosity
distance, the observer has to live in a bubble with a larger expansion rate.
In this setup, the average over all the phenomenological observers gives the
global observer who sees the EdS luminosity distance, but, again, the variance
is too big for the average to be meaningful. In other words, even if the GBS
describes perturbatively the spacetime, it does not give the PBS of the
phenomenological observers. 

\section{An observable backreaction}

As a final point, we now focus on the PBS which encapsulates
the desideratum to be tied to observational data, that is, to fit observations
by a model \cite{Ellis:1987zz,Marra:2007gc,Buchert:2002ht,Buchert:2002ij,Hellaby:1988zz}.
We stress again that the ABS, instead, is not directly related to observations.

Fitting along the light-cone is actually what is done by the concordance
$\Lambda$CDM model, which happens to be a good fit to observational data. We
therefore define:

{\bf Observable Backreaction:} the effect of the evolution of inhomogeneities
that leads the PBS to have an expansion history and curvature different
from those deduced from the corresponding local quantities.

The idea is that the former is close to that of a $\Lambda$CDM model whereas
locally we are dealing with a dust model.
Moreover, the trigger is the evolution of large-scale structures which became
nonlinear recently, exactly when a primary dark energy is supposed to start
dominating the energy content of the universe and causing acceleration. The
coincidence problem is not, therefore, a problem anymore, but a trigger event.

We can now understand why we used the term ``backreaction'' for the weak
backreaction while usually it is reserved for strong backreaction only.
Indeed, as discussed in the examples of the previous Section, it is possible that weak backreaction
can observationally mimic strong backreaction, in the sense that both can give PBS qualitatively different from the GBS.
Therefore, it might not be possible to distinguish between weak and strong backreaction and hence
the reason to focus only on observable backreaction.

Summarizing, the observable backreaction is only concerned with the ``end
result,'' which is the only result physically meaningful: its definition, indeed, does not mention GBS or ABS.
The distinction made by introducing strong backreaction
\cite{Buchert:2007ik,Wiltshire:2007jk,Rasanen:2006kp,Kolb:2005da,Coley:2005ei,Behrend:2007mf,Brown:2009cy,Larena:2008be,Clarkson:2009hr,Bolejko:2008yj,Kai:2006ws,Paranjape:2007wr}
and weak backreaction
\cite{Marra:2007pm,Marra:2007gc,Alnes:2005rw,Valkenburg:2009iw,Mustapha:1998jb,Celerier:1999hp,Tomita:2000jj,Iguchi:2001sq,Chung:2006xh,Enqvist:2006cg,Alexander:2007xx,Brouzakis:2008uw,GarciaBellido:2008yq,Zibin:2008vk}
is indeed useful to lay a framework and separate the possible sources of
backreaction we have in a model, but we need to keep in mind that only the PBS
matters. If there is no dark energy, indeed, we should not, and maybe we can
not, be concerned with which of the backreactions is responsible for the
observations.

Finally, we point out that, even if the backreaction proposal is not able to
explain away dark energy and an other physical cause is needed, it still
is of interest within the precision cosmology we will be facing in the future.

\begin{acknowledgements}
It is a pleasure to thank Marie-No\"elle C\'el\'erier and Simon DeDeo for useful discussions.
This work was supported in part by the Department of Energy. V.M.\ would like
to acknowledge the Kavli Institute for Cosmological Physics for hosting a visit
to the University of Chicago and ``Fondazione Angelo Della Riccia'' for
support. S.M.\ and V.M.\ acknowledge ASI contract I/016/07/0 ``COFIS" for
partial financial support. 
\end{acknowledgements}



\end{document}